\documentclass[11pt]{article}
\usepackage{preamble}
\usepackage{hyperref}
\usepackage{csquotes}
\usepackage{nameref}
\usepackage{tiffany}
\def\beq{\begin{equation}}
\def\eeq#1{\label{#1}\end{equation}}
\def\eeqn{\end{equation}}

\newenvironment{Eqnarray}%
   {\arraycolsep 0.14em\begin{eqnarray}}{\end{eqnarray}}
\def\beqa{\begin{Eqnarray}}
\def\eeqa#1{\label{#1}\end{Eqnarray}}
\def\eeqan{\end{Eqnarray}}

\let\bar=\overbar

\def\lsim{\mathrel{\raise.3ex\hbox{$<$\kern-.75em\lower1ex\hbox{$\sim$}}}}
\def\gsim{\mathrel{\raise.3ex\hbox{$>$\kern-.75em\lower1ex\hbox{$\sim$}}}}

\def\del{\partial}
\def\Dslash{\not{\hbox{\kern-4pt $D$}}}
\def\dslash{\not{\hbox{\kern-2pt $\del$}}}
\def\pslash{\not{\hbox{\kern-2pt $p$}}}
\def\ETmiss{\not{\hbox{\kern-4pt $E$}}_T}

\def\Dlr{\mathrel{\raise1.5ex\hbox{$\leftrightarrow$\kern-1em\lower1.5ex\hbox{$D$}}}}

\def\MSB{{\bar{M \kern -2pt S}}}
\def\msb{{\bar{\scriptsize M \kern -1pt S}}}

\def\drb{{\bar{\scriptsize D \kern -1pt R}}}

\author[1]{Matt Bellis\orcidlink{0000-0002-6353-6043}}
\author[2]{Brian Shuve}
\author[2]{Anna Barth}
\author[2]{Andr\'es Cook}

\affil[1]{Siena College, 515 Loudon Road, Loudonville, NY 12211-1462, United States}
\affil[2]{Harvey Mudd College, 301 Platt Blvd., Claremont, CA 91711, United States}

\title{Opportunities for theory studies with public collider data}

\begin{document}

\date{\vspace{-5ex}}
\maketitle

\medskip

 \begin{abstract}
     \noindent Over the last 20+ years, experimentalists have presented tantalizing hints of physics beyond the standard model, but nothing definitive. With the wealth of data from experiments, in particular the collider experiments, it is imperative that the community leave no reasonable model untested and no search unsought. Open datasets from particle physics experiments provide a relatively new and exciting opportunity to extend the reach of these searches by bringing in additional personpower in the form of the theory community. Analysis of these datasets also provides the opportunity for an increased information flow between theorists and experimentalists, an activity which can only benefit the entire field. This paper discusses the potential of this effort, informed by the successes of the last 5 years in the form of results produced by theorists making use of open collider data, primarily the datasets released by the CMS collaboration. Concerns about the potential negative impact on the field are also discussed. For a more detailed accounting of these concerns, see Ref.~\cite{Nachman2022}.
\end{abstract}

\newpage
\tableofcontents 
\clearpage

\def\thefootnote{\fnsymbol{footnote}}
\setcounter{footnote}{0}

\section{Introduction}

In January 2009, members of a variety of particle physics experiments came together at DESY for the first
DPHEP (Data Preservation in High Energy Physics), followed by a second workshop at SLAC in May of 2009
and a third workshop at CERN in December 2009. Some experiments had been shuttered only recently 
(prmarily at the HERA (DESY) and PEP-II (SLAC) colliders) and wanted to preserve their data for future access while other experiments
were just starting to take data (CMS) and wanted to begin the discussion early about how to keep the data
accessible for as long as possible. In an influential white paper~\cite{dphepWP}, the group identified
four (4) levels of open data access, the two highest levels of which would allow
for publication-quality results to still be produced. 

\begin{table}[]
    \centering
        \caption{Various preservation models, listed in order of increasing complexity. Subsequent
models are inclusive. For example, preservation model 4 also includes steps and use cases
described in models 1,2, and 3. (Table from ~\cite{dphepWP})\label{tab:dphep} }
    \begin{tabular}{p{3in}|p{3in}}
    \hline
{\bf Preservation Model} & {\bf  Use case} \\
\hline
1. Provide additional documentation & Publication-related information search \\
2. Preserve the data in a simplified format & Outreach, simple training analyses \\
3. Preserve the analysis level software and data format & Full scientific analysis based on existing 
reconstruction \\
4. Preserve the reconstruction and simulation software and basic level data & Full potential of the experimental data \\
    \end{tabular}
\end{table}

These workshops were inspired in part by an analysis by members of the JADE collaboration~\cite{Bethke:2009ehn}. 
The JADE experiment recorded data produced by collisions from the PETRA $e^+e^-$ collider at DESY from 
1979-1986. In 2008, collaborators reanalyzed the data making use of new next-to-next-to-leading-order (NNLO)
theoretical calculations and extracted an improved measurement of the strong coupling constant $\alpha_S$.
This use of older data with new theoretical input generated some excitement at the possibility of 
unearthing hidden scientific
gems in older datasets as well as a sense of purpose that the community should put effort and
resources into preserving
these older datasets {\it and} preparing the newer datasets for an eventual time when they could be archived
but still accessible and analyzable.

If one reads the white paper~\cite{dphepWP} from those early workshops one finds that the focus is on 
collaboration members, or other experimentalists, analyzing their collaboration's older data in light of new theoretical ideas or improved understandings
of backgrounds. Nowhere in the paper does it explore the idea of theorists or phenomenologists 
(or non-HEP scientists) performing
their own analyses with these datasets. And yet, since 2017, a significant amount of new measurements
coming from the LHC open datasets have been from theorists. The lack of experimentalists publishing
with these datasets can be partly attributed to the fact that the LHC experiments are still taking data,
but that should not take away from the efforts of the theory community to exploit these data. 
In this paper, we summarize some of these analysis, discuss the experiences of those theory analysts, 
point out the future opportunities, and also discuss some of the social barriers that should be addressed
when it comes to the person power allocated to maintaining these datasets. This paper focuses on collider experiments, particularly the LHC experiments with an emphasis on the CMS open data, the largest open dataset of experimental collider data to date. 
A very significant anecdote regarding correlation functions in jet substructure is also discussed as it 
demonstrates potentially the most important outcome of the open data effort which is the increased
flow of ideas between the theory and experimental communities.

\section{Open data from the LHC experiments}

At the time of this paper, all four (4) major LHC experiments (ALICE, ATLAS, CMS, LHCb) as well as the 
OPERA neutrino experiment and the PHENIX (at RHIC), have contributed data to the CERN Open Data Portal~\cite{CODP}. 
Not all of these data are intended to be used for publishable analysis but instead are limited
to skimmed data for outreach or specialized studies (e.g. jet reconstruction training~\cite{https://doi.org/10.7483/opendata.atlas.l806.5cku}). 
On the other hand, CMS has released almost 2 petabytes
of collision data and Monte Carlo data and this dataset has generated the majority of new
publications and measurements. For this reason, much of this paper will focus on the use of CMS data.
The lessons learned from the CMS experiences are applicable to all collider experiments
and it is anticipated that more data will be available from other LHC experiments
in the future. Details of these datasets and the policies regarding their use can be found
from links on the Open Data Portal or summarized in some helpful review articles, 
for example Chapter 4 in ``Unveiling Hidden Physics at the LHC"~\cite{Fischer:2021sqw}.

\section{Studies to-date}

\subsection{CMS open data} 

Since 2017, there have been a number of published analyses (15-25) from non-CMS groups using CMS open data,
some of which are found in the bibliography. Perhaps the most significant publication
is the first from the MIT group led by Jesse Thaler, 
``{\it Jet Substructure Studies with CMS Open Data}"~\cite{Tripathee:2017ybi}. The impact stems not just
from the 
physics content but for
demonstrating to the community that this work {\it can} be done by theorists, that there is a desire to do this work, 
{\it and} for providing feedback on their experience in the paper. Section V of that paper is titled ``Advice to the Community"
and contains subsections ``Challenges" and ``Recommendations", both of which had an seismic effect on the CMS Data Preservation and Open Data (DPOA)
Working Group, leading to a series of workshops starting in 2020 at which participants were led through the 
process of accessing and analyzing the open data. This paper was quickly followed by another
from the same group~\cite{Larkoski:2017bvj}, also analyzing jets. This effort developed a significant amount
of expertise within the Thaler group, allowing for more efficient work with the open data, 
some of which will be discussed in the next section.

In 2019, a very different analysis was performed by Cesarotti et al, ``{\it Searching in CMS Open Data for Dimuon Resonances with Substantial Transverse Momentum}"~\cite{Cesarotti:2019nax}, less focused on jet substructure than a search for new
particles. At the same time, two of the authors, Jesse Thaler and Matt Strassler, had a Letter to the Editor published
in Nature ``{\it Slow and Steady}", mentioning their experience using the data for this search, but also making a real
call for more open data, in part because it can allow for ``testing unconventional strategies and potentially leading to unexpected discoveries, new approaches and fruitful discussions". 

It is difficult to keep track of exactly how many publications have come out at any given moment from
the CMS open dataset or the the other LHC
experiments' open data. Different search criteria on iNSPIRE yield 59 or 25 papers~\cite{inspire:cod1, inspire:cod2}, depending
on your search criteria. While some of the papers are talking {\it about} open data, rather than using it, there is an
impressive number of publications from the various groups, not all of whom are theorists.
A subset of these papers can be found in the bibliography~\cite{Tripathee:2017ybi, Larkoski:2017bvj, Knapp:2020dde, Andrews:2018nwy, Komiske:2019jim, Andrews:2019faz, Lester:2019bso, Komiske:2022enw, Cesarotti:2019nax, Apyan:2019ybx, PaktinatMehdiabadi:2019ujl}.

For any non-CMS scientist looking to analyze the CMS data, the appendices in the first paper from Thaler et al\cite{Tripathee:2017ybi} 
about their experience and suggestions is a must read. In a similar vein, we include in this whitepaper an accounting
from Brian Shuve (a co-author of this paper and faculty member at Harvey Mudd college) about his and his students'
experience with CMS open data in Appendix~\ref{sec:app_hm}. Their work did not yield a new publication, which makes
it particularly valuable as it gives insight as to what might be holding other groups back from 
producing new measurements. Shuve also provides some commentary on the use of open data by smaller institutions
with fewer resources and the benefit students get from exposure to this type of analysis.

At the early DPHEP meetings, theorists were not considered as potential users of these data. Even after the initial 
release of the CMS open data, it was unclear as to whether or not particle physics data was inherently too
complicated to be analyzed by scientists outside of the experimental collaboration who recorded the data. However, 
the past 5 years have shown this is not true. While the data {\it is} complicated, concerted efforts by the 
experimental groups, particularly CMS, have produced enough documentation that these data can be used
by external scientists. Within CMS, these publications and related workshops have provided concrete feedback
on how the data can be made more easily accessed and analyzed.
This growing community of theorists who have wrestled with the data and the messiness of applying corrections and systematic
studies and the like, will inevitably spawn novel ideas and approaches to teasing signals out of the growing LHC datasets.
It is with this goal in mind, that the CMS DPOA group has committed time and resources to efforts like the CMS Open Data Workshops~\cite{odworkshop} and the CMS Open Data 
Guide~\cite{odguide} that work to lower the threshold for new users to access and analyze these data,
not just for theorists but for any interested scientist.

\subsection{Other experiments}

It is not only CMS that has released data for public access, though the working model can be different
and it is worth highlighting these other approaches. The ALEPH collaboration archived $e^+e^-$ 
data from LEP1 (91 GeV) and LEP2 (91-209 GeV). 
Whereas CMS has provided data and documentation through the CERN Open Data Portal, this work
proceeded differently with ALEPH physicists working closely with non-ALEPH scientists. 
Using these data a collection of scientists published or are working on the following:
\begin{itemize}[itemsep=0pt,parsep=0pt,topsep=0pt,partopsep=0pt]
    \item First search for collective effects in $e^+e^-$ collisions using two particle 
    correlations~\cite{Badea:2019vey}
    \item First measurement of anti-$kT$ jet-energy spectra and substructures compared to various event generators, NLO, and NLL'+R resummation calculations~\cite{Chen:2021uws}
    \item Application of novel machine learning method Omnifold to perform unbinned unfolding~\cite{ichep_badea}
\end{itemize}
This approach where the experimentalists 
provide perspective and feedback while an external group (theorists or otherwise) performs much of the analysis, 
is an interesting approach and one perhaps best suited to older experiments where there is not
the resources anymore to develop documentation and infrastructure to allow others to freely access the data.
It does rely on the experimental collaboration members doing some ``volunteer" work to interface with external users, in the event
there is no official support for these efforts.

\section{Opportunities in the future}

At the time of this paper, the LHC experiments are about to embark on the Run 3 data-taking period and it is worth considering
what can be done with these upcoming data, in light of past successes and the challenges ahead. 

\subsection{Correlation functions and a story}

Let us start with an anecdote relayed by Ian Moult (Yale) at KITP (Kavli Institute for Theoretical Physics) at 
a pre-Snowmass workshop Feb 23-25, 2022\cite{MoultKITP22}. Moult discussed a previous workshop at KITP in 2009 and an  exchange between Juan Maldecena and Joseph Polchinski about 2-point correlation functions (2PCFs) in jet final-state particles. In 2008, Maldecena and his collaborator, Diego Hofman, had published an article looking at how some conformal field theories could be tested by measuring at these 2PCFs in jets\cite{Hofman:2008ar}. At the 2009 KITP meeting, it was asked (paraphrased by Moult) why experimentalists hadn't done this yet and the impression was that the experimentalists had just not thought about this approach much at the time. 

Experimentalists parameterize jets in terms of {\it shapes}, an approach developed in the 1970's and 1980's, which worked well for the experimental analyses but did not directly map on to the field theories. This direct connection between theory and observables is what most scientists want, theorist
and experimentalist alike. The work from Maldecena and others demonstrated that there should be a scaling behavior in these observables, however to see this, the higher energies of the LHC were needed, as opposed to previous $e^+e^-$ experiments. 

By 2019, Jesse Thaler had built up expertise and infrastructure to work with open data within his group. 
Building on their experience from jet substructure and in collaboration with Moult, they were able to directly
compare QCD theory to the data using these correlation functions\cite{Komiske:2022enw}.
Similar analyses being done with higher-order correlators, allowing for a deeper exploration of how well these
theories predict these observables. The theory community can now explore jet structure using a language
tied to the deeper, more fundamental theory.

It could be argued that there was no need for the theorists to perform these analyses and that these correlation functions could have been measured by the experimentalists at CMS or ATLAS. But the fact
remains that the experimentalists did not prioritize these particular measurements. That doesn't mean
that people didn't think they were important measurements to make! Instead, it highlights that if there
is finite available personpower which is divided between a wide range of searches as
well as calibrations and service tasks, there is a danger of interesting tests slipping through the cracks. 
We all have our own biases as to what is important to work on and it is almost impossible for us to know in advance
where is the {\it right} place to put our efforts. But to one of the experimentalist authors on this paper (Bellis), this 
is a profound story about how open data lets the theory community tests things directly. There are any 
number of reasons why these correlation functions were not measured previously, 
and it will be very interesting to see how
this work progresses and what insight to we gain in our understand of QCD (theory) or even perhaps how to 
reconstruct and parametrize jets in our detectors (experiment).

\subsection{Additional hopes and musings}

It is almost impossible to say with certainty what the theory community can do with the open datasets from the
collider experiments. If it was obvious it would have been done by now. In preparing for this paper, there
were some very useful brainstorming sessions with the Snowmass TF07 conveners, Fabio Maltoni (Università di Bologna), 
Shufang Su (University of Arizona), and Jesse Thaler (MIT), all theorists. 
This section attempts to capture those discussions and the optimism about future work.

An topic that kept coming up is that open data allows theorists to do exploratory studies themselves, without
taking up time and personpower from the experimental collaborations. Imagine that a scientist develops
a rather baroque model, one that is consistent with previous observations, and is not sure if a 
search is even worth doing with current experiments. 
With the open data, often a subset of the larger dataset, the theorist can run their
own preliminary study. It may even be that the actual collision data is not necessary
and that the Monte Carlo is enough to make a statement about
the likelihood of observing a signal. Even if a new published measurement
does not come out of that study, it might be enough to encourage the experimentalists to perform the study themselves
with the full datasets...or maybe it is not worth doing. In either case, valuable knowledge has been gained and the
theorist benefits from the increased agency they have in directing experimental efforts. In the end, this might
wind up saving the experimental collaboration time by keeping them from going down dead ends, thanks
to the preliminary work of the theorists.

It should be noted that some would say that in the platonic ideal, a theorist would simply go to the experimentalists
and ask them to check their theory. But this denies the reality in which we live in which experimental groups have
finite resources and need to make decisions about where to concentrate their efforts. This can exclude potentially
exciting work, not through malice, but just recognizing that different people have different priorities.

One limitation of this approach is that there might not exist the {\it right} Monte Carlo to perform these studies.
In that case, it is essential that the collaborations make available the software to simulate any of these
new proposed physics processes. But this takes additional personpower. 

Beyond any one person's pet theory, it should be obvious that we are in a different place as a field than we
were 10 years ago. The Higgs was discovered, but except for new hadronic states, there has not been much else
beyond some tantalizing tensions (e.g. lepton-universality violation?). We need different perspectives on how 
to attack these still-unanswered questions (matter-antimatter asymmetry, dark matter, dark energy,
mass hierarchy, etc.). The involvement of more theorists looking at the ``raw" data should lead 
to new perspectives on how to analyze these data or how to include these measurements in global fits.
Everyone of us in the field who received a PhD in Physics took what were effectively {\it theory}
courses in graduate school and this benefited experimentalists by providing some understanding of what
we were looking for. It seems obvious that theorists can also benefit by deeply engaging with 
open datasets and learning at a deeper and more intuitive level {\it how} these analyses are performed. 

On a more practical level, the more outside members make use of the open datasets, whether theorists or not, 
the more checks we will have on those data and the tools that people are using to access and analyze them. 
One can optimistically imagine a future with an active and interconnected community of open data users,
all of whom are helping each other solve problems in the same way that internal member of CMS (for example)
work together. 

Continuing in that practical vein, from a social engineering standpoint, the more theorists who make use of
these data, the more will {\it want} to use it for their own purposes. The shared experience of the community 
will inevitably make it easier to onboard new theory analysts. And if there are continued successes 
(publications) that
come out of this theory/open-data effort, it can provide a road map for other experimental fields
(e.g. dark matter direct detection) to perhaps release their data.

\section{Concerns about open data}

Since the early DPHEP meetings, concerns have been raised about whether or not it is worth it to make collider data open. Some of the earliest fears revolved around well-meaning, capable non-experts who would mistakenly find signatures in the data that suggested New Physics, requiring the original experimental collaboration to go and ``put out fires" and potentially causing a drain on people's time. This concern can be somewhat alleviated with better documentation and more standardized user tools that cut down on the likelihood of simple mistakes and to date, this situation has not manifested. There have been no claims on BSM (Beyond the Standard Model) physics, coming from non-members analyzing CMS data, or any other open collider data set, to the authors' knowledge.

Another valid concern that is difficult to quantify, is simply the question ``is this worth it"? Does it benefit a collaboration to divert resources and person power to making their datasets available to theorists or other experimentalists, when those finite resources and time could be put to better use within the collaboration. 
This is a tough question, and there is no there is no universally obvious solution to these concerns. 

The astrophysics community has a culture of open data, with high-profile missions 
like Gaia and Fermi making their data
publicly available to anyone who wishes to use it. Early concerns had been raised 
in that community about scientists
who built, launched, and then calibrated the data getting scooped once the data was released, in part
because they  had done all the heavy lifting and had not yet had time to develop their own analysis pipelines.
This concern
was address by implementing embargo periods (ranging from 6 to 12 months) 
when the data was {\it only} accessible
to the mission's collaborators for some period of time after calibration. This gave the collaboration 
first shot at any significant discoveries. A similar embargo period has been proposed by some whereby the 
data is not released until 5 years after data taking, under the argument that 5 years is enough time for 
the experimental collaboration to make most of the important measurements. 
However, given the challenges and demands
on graduate student and post-doc time to do calibrations and data quality checks, maybe 5 years is not enough.
In any event, the experimental community should take a hard look at the 
cost-benefit for experimental collaborators
and where they put their efforts, all of which impact their getting a 
position at the next stage of their careers. 

There are no easy answers and the hope is that this will remain a topic of discussion going forward. 

\section{Summary and recommendations for the experiment and theory communities}

The release of open data, particularly from the CMS experiment, has been a unqualified success. 
The results from jet substructure and correlation functions are sure to encourage new approaches and 
discussions between the theory and experimental communities. 
The demonstration that theorists can perform new searches for resonances without waiting for 
experimentalists should both excite and motivate both communities to do more, both with the open data and
within the collaborations with the full datasets. With that in mind, we offer some minimal 
suggestions. 

\noindent{\bf Lower the threshold for access and analysis}.
Going forward, we would encourage the experimental collaborations to continue efforts to make 
analysis-quality data {\it and} code available, as well as developing documentation and simplified
workflows for accessing these data. Similarly, the theory community has an opportunity to 
gather their own resources and perhaps develop their own training materials and workshops, 
in collaboration with the experimentalists. It is imperative that both groups engage in active discussions
about these activities so that effort is not wasted. Without proper feedback, the experimental groups
have no guide as to what the theory community actually wants or needs. Simplifying the access 
to the {\it simulation software} would be a big help for theorists who want to test their models
themselves. If the field dedicates real resources to this now, what will be possible in 10 years? 20 years?
Much of the tools exist so maybe the answer is simply better training. 

\noindent{\bf Allocate resources}.
For the community to develop these new tools and new approaches, 
resources must be provided. These resources, for example to build new tools to simplify access, 
need to come in both the form of funding to pay people to build these tools {\it and} in valuing 
this work enough to assign people's time and efforts to these projects. The support from both 
CERN and CMS has been truly impressive so far, but it will only continue if there is a steady stream
of funding support. One model is for {\it theorists} to apply for funding to work not just
on analysis of these data, but to contribute to the documentation and computational tools
that are needed to access the datasets.

\noindent{\bf Continue to evaluate the incentives and workflows for experimentalists.}
Finally, it is worth reiterating the concerns that putting effort into producing and maintaining these
datasets might not be in the best interest of the experimental collaboration or individual 
experimentalists and their career paths. Since some of the concerns are correlated with the time
it takes to complete a full analysis, perhaps this motivates the experiments to change their workflow and 
training of their own graduate students. Perhaps there are often-repeated tasks (e.g. initial skims of the data)
which should be offloaded from grad students and replaced with a service model where analysts
``place an order" for some subset of collision data and Monte Carlo samples with some cuts, freeing themselves
up to focus on the physics and innovative approaches to inferring the underlying processes.
This is fairly speculative, but worth further discussions.

For many experimentalists, much of our excitement comes from the {\it potential} of 
any dataset and what secrets it holds. 
It is obvious that this excitement is shared with our theory colleagues who are 
discovering the joys and heartaches
of wrestling with real data. We are optimistic about the open data efforts 
and look forward to see what fruit they bear in the years to come. 

\subsection*{Acknowledgements}
Matt Bellis would like to thank Fabio Maltoni, Shufang Su, and Jesse Thaler, for initiating this white paper and
for useful discussions during the writing. He would also like to thank the broader CMS collaboration for their commitment to data preservation and open access and in particular CMS members
Kati Lasilla-Perini, Edgar Carerra, and Clemens Lange for their work and championing of these efforts. Perspective has also been provided by Ben Nachman and this
paper should be considered in light of more detailed concerns and views he has outlined in 
a related whitepaper~\cite{Nachman2022}.
We also thank Anthony Badea for drawing our attention to the 
recent and ongoing work with the ALEPH dataset.
Contribution by Matt Bellis is based upon work supported by the 
National Science Foundation under Grant No. PHY-1913923.
\clearpage

\bibliographystyle{utphysmod}
\bibliography{main.bib}
\clearpage

\appendix 
\section{Experience with CMS Open Data at Harvey Mudd College}\label{sec:app_hm}

This section is a detailed account of the work by a group at Harvey Mudd, led by Brian Shuve, using 
CMS open data for a new search. The text that follows was written by Shuve (one of the co-authors of
this whitepaper) and is important for the immediate perspective of a theorist working with these
open datasets. While the attempt at a new result did not yield a publication, valuable experience
was gained by the students.

It should be noted that Harvey Mudd is an undergraduate-only college and so while the group 
had access to helpful and supportive CMS members, the bulk of the work was done by Harvey Mudd
undergraduate students. Thus their experience may differ from a group with more experienced
graduate students and post-docs, not so much for their advanced education, but for the amount of time
they have to dedicate to an analysis. Even the most passionate undergraduate research experiences are often severely limited
as their work tends to be done over summers or as 1-3 credit independent studies.

\subsection{Goals}
The primary purpose of our work was to use CMS Open Data to investigate a hidden-sector signature that had not yet been the target of an official CMS or ATLAS search at the time. This would allow us to assess the viability of the search strategy, understand the backgrounds and compare to Monte Carlo simulations, and potentially to set constraints on new hidden-sector interactions. The particular model we were investigating was a dark Higgs-dark photon model proposed in Ref.~\cite{Blinov:2017dtk} motivated by the dark Higgs-strahlung process, $pp\to Z\to A' h_{\rm D},\,h_{\rm D}\to A'A'$, where $A'$ is a dark photon and $h_{\rm D}$ a dark Higgs boson. With the dark photon decaying leptonically, this gives rise to a 4-lepton resonance from the decay of the dark Higgs.

As a secondary goal, we were interested in assessing the viability of using CMS Open Data as a research tool in primarily undergraduate institutions (PUIs), particularly those that do not have faculty that are members of major high-energy experimental collaborations. Open data is potentially a game-changer in terms of access to resources and the ability to do research with experimental data in high-energy physics, both of which are typically lacking at most PUIs. Our team was composed of the PI and two rising sophomore summer student researcher.

At the time, access to CMS Open Data was via a downloadable Virtual Machine (VM) image with the CMS environment pre-installed. The available data at the time was primarily from the 2010--2012 datasets of Run 1.

\subsection{Team and Timeline}
The work was undertaken in Summer 2018. Our team consisted of the PI (Brian Shuve), who acted primarily in an advisory capacity, and two rising sophomore students (Anna Barth and Andr\'es Cook) who did the bulk of the work. The students had no prior experience with research in high-energy physics, while the PI had  experience with data analysis as a member of the BABAR Collaboration but no such experience with the LHC experiments.

\subsection{Analysis}
The first step was to learn the basics of the CMS environment and ROOT fitting, including processing events and plotting histograms. We did this using the PATtuples ({\it a particular data format used by CMS}) provided in the 2011 CMS Outreach Exercise, which includes events collected with dilepton triggers. The provided PATtuples and software allowed us to very quickly ``rediscover'' the $Z$ and various meson resonances, as well to investigate the $Z$ width, and lepton multiplicities and kinematics in these events.  This allowed us to explore, for example, the number of events with very high multiplicities of leptons ($>4$), which would otherwise be unavailable to theorists in the absence of Open Data.

Since our signal consists of a 4-lepton resonance, our next task was to observe and estimate the $Z\to 4\ell$ branching fraction following the selections in Ref.~\cite{CMS:2012bw}. We observed a peak at the $Z$ mass and estimated the fiducial cross section for $Z\to 4\ell$, finding agreement with the result from Ref.~\cite{CMS:2012bw} but with larger uncertainties because of the more limited public dataset.

Finally, we proceeded to investigate 4-lepton resonances at other masses. The lack of signal Monte Carlo (MC) with the proper detector simulation made it challenging to assess both the signal efficiency and resolution. To investigate the efficiency, we first modified the code to generate PATtuples from AOD format ({\it a different data format used within CMS and which is used for the bulk of the 2011 and 2012 CMS open data}) and used it to generate PATtuples from the AOD files of Drell-Yan MC and Higgs decays to 4 leptons at several masses. From this, we could estimate lepton efficiencies as functions of $p_{\rm T}$ and $\eta$, and then apply the efficiencies to signal events generated in MadGraph 5 \cite{Alwall:2011uj,Alwall:2014hca}. Because of the relatively large uncertainties in the extracted efficiencies per lepton, there were very large uncertainties in the overall signal efficiency when propagated to the 4-lepton final state, making it challenging to go from a fiducial cross section limit to a limit on the dark Higgs production cross section. We also had difficulty assessing trigger efficiencies. When applied to the measurement of the $Z\to4 \ell$ branching fraction, our method agreed with the result in Ref.~\cite{CMS:2012bw}, but the uncertainties were very large due to uncertainties in the efficiencies.

As a final step, we implemented a crude method to extract a mass-dependent signal significance for the dark Higgs fiducial cross section:~we modelled the background over short $4\ell$ mass intervals as a quadratic function, used this to estimate a number of background events in a bin equal in size to the signal resolution, and compared to the observed number of events in that bin. We then used a likelihood method to set approximate upper limits on the number of signal events (after selections) consistent with the data. We did not use a more sophisticated method, in part because of  time constraints, and also given the large uncertainties in the efficiency it was in any case difficult to obtain meaningful limits on the dark Higgs model. With more time, we could have developed a more sophisticated fits and statistical analysis using the  ROOT tools embedded in the CMS environment.

Beyond the physics analysis, we developed a pipeline for converting AOD files to PATtuples and further compressing these to CSV files, allowing for relatively efficient tweaks to the analysis as we went.

\subsection{Successes and Challenges}
In some ways, it was remarkable what the group accomplished over the span of only 10 weeks. Within a week, we were able to use existing PATtuples to extend existing analyses in new directions, allowing us to better understand the kinematics and rate of multilepton events. It was also relatively straightforward to use the provided code to convert AOD files to PATtuples allowing us, for example, to quickly generate new PATtuples of MC samples that were not part of the 2011 Outreach Exercise datasets. The installation of the VM was straightforward, and the use of pyROOT allowed us to quickly implement selections, plotting, and fits to the data. CMS Open Data therefore provides a unique opportunity for undergraduate students and other interested non-experts to quickly engage with real LHC data and learn physics from the publicly available samples.

At the same time, we faced multiple challenges, most of which were related to a lack of documentation and support for the use of Open Data. For example, when generating PATtuples from the AOD MC samples, we found it straightforward to access the reconstruction of the events but not the simulated information. At one point, members of our group were reduced to guessing the names of tags that we hoped corresponded to generator-level information; remarkably, we succeeded in this attempt, but it indicates the somewhat random nature of trying to access important information without documentation or collaboration members to talk to. We also ran into issues assessing trigger efficiencies; we were able to access stored trigger objects when converting AOD files into PATtuples, but all of the objects were the same for every event and did not seem to actually contain information about whether the event passed the trigger (or which triggers were activated). In many cases, when we attempted to access documentation, we were directed to parts of the collaboration website that were restricted. We had some success with asking friends who happened to be CMS collaboration members for help, who  in turn directed us to relevant individuals within CMS, but we ran out of time before we could thoroughly resolve the issue.

Since we were running the CMS environment VM on personal computers, we were also limited in terms of computing resources. A ready interface that allows analyses to be scaled up on high-performance computing or cloud computing systems would be helpful.

Perhaps the most significant obstruction to completing the analysis was the lack of signal simulation, which inhibited estimations of the signal efficiency and resolution. These are both crucial to obtaining meaningful results  applied to the dark Higgs model. While we could obtain somewhat reasonable estimates by extracting signal efficiencies and resolutions from Higgs to 4 lepton events, this procedure can run into problems for signal masses far from those used in the simulations; this problem is exacerbated now that we actually know the SM Higgs mass, and presumably there are fewer such simulations at different masses being produced for Run 2.

\subsection{Status of Project}

At the end of the summer research session, we assessed our progress and the obstacles we would need to overcome to obtain a result that could be published in a peer-reviewed journal. Ultimately, we felt that with the resources available at an undergraduate college (in terms of research time as well as access to CMS collaboration members to help us overcome technical challenges), we did not see a viable path forward, especially with experimentalists within the collaborations working on versions of this search. Since our group is also a part of the BABAR Collaboration, we pivoted the experimental part of our group to working on analyses  with BABAR data.

Since our most recent experience with CMS Open Data was in Summer 2018, it is possible that recent developments have overcome or mitigated some of the challenges we faced as outlined here.

\subsection{Opportunities for future studies and perspectives}

Open data can potentially expand the range of investigations performed using LHC (and other collider) data. Particularly for models of dark matter and hidden sectors, for which there exists a vast range of signals compared to simpler weakly interacting massive particle (WIMP) scenarios, there will always be more search proposals than capacity for doing analyses within the collaborations. Open data can provide opportunities for theorists or experimentalists outside of the collaboration to perform analyses and better understand backgrounds; it can also provide an important legacy for the experiment such that analysis can continue after the collaboration has ceased activity. Examples include various combinations of resonances and associated objects (such as jets, leptons, and missing momentum), displaced vertices, and other signatures that can be reconstructed with standard analysis tools.

At the same time, all of this is possible only if tools and documentation are provided that allow for this type of us of open data. In our experience, this means providing better documentation of how to access information from the raw data files, software for performing simulation of novel signals (even if it is not quite at the level of the full internal detector simulations), and potentially better tools for streamlining analysis and fitting. We found that it was relatively straightforward  to develop tools based off of examples provided by the collaboration, and much more challenging when we needed to do anything that was not already present in the public tools. It might therefore be helpful to have a public repository of versions of existing analyses by the collaborations that have been ported over to the open dataset which can be used by the public as a basis from which to work. It would also be beneficial if there existed over the next 5--10 years specific point people within the collaborations that can work with users of open data to understand what is currently working well and what is missing, as this can better inform the experiments about how access to open data can be improved in the future. Finally, more collaboration and sharing of tools between users of open data could be helpful; for example, early studies by the MIT Open Data group led to the development of publicly accessible tools that are useful for all subsequent users of open data \cite{Tripathee:2017ybi,Larkoski:2017bvj}. A robust and active community of open data users will allow for support, encouragement, and constructive criticism for participants' work.

Finally, open data can serve an important purpose in granting access to high-energy experimental data at institutions without dedicated particle experimental groups. While outreach is a part of the possible uses of open data, research-quality work can also in principle be done with open data at primarily undergraduate institutions (PUIs), historically Black colleges and universities (HBCUs), and other minority-serving institutions (MSIs) that might not otherwise have access to this type of data. To facilitate this type of access, partnerships between the collaborations and PUIs, HBCUs, and MSIs could provide the training and support needed for students and faculty at these institutions to use open data. It would also be helpful to provide tools that allow for interface with local high-performance facilities or cloud computing, which is needed for large-scale analyses.

\vspace{1.0cm}

\noindent \emph{Acknowledgements:}~We (the Harvey Mudd group) are grateful to Yangyang Cheng for helpful conversations during the course of this project.


\end{document}